\documentclass[twocolumn,showpacs,prl]{revtex4}
\usepackage{graphicx}%
\usepackage{dcolumn}

\begin{document}

\preprint{HEP/123-qed}

\title{Observation of Quantum Hall Valley Skyrmions}
\author{Y. P. Shkolnikov, S. Misra, N. C. Bishop, E. P. De Poortere, and M. Shayegan}
\affiliation{Department of Electrical Engineering, Princeton University,
Princeton, New Jersey 08544}

\date{\today}

\begin{abstract}
We report measurements of the interaction-induced quantum Hall
effect in a spin-polarized AlAs two-dimensional electron system
where the electrons occupy two in-plane conduction band valleys.
Via the application of in-plane strain, we tune the energies of
these valleys and measure the energy gap of the quantum Hall state
at filling factor $\nu$ = 1. The gap has a finite value even at
zero strain and, with strain, rises much faster than expected from
a single-particle picture, suggesting that the lowest energy
charged excitations at $\nu=1$ are "valley Skyrmions".
\end{abstract}


\maketitle

In a two-dimensional electron system (2DES) with a spin SU(2)
symmetry, electron-electron interaction causes a dramatic
deviation from the single-particle physics. For example, the large
Coulomb energy cost of a spin flip results in a large excitation
gap for the $\nu=1$ quantum Hall state (QHS) even in the limit of
zero Lande
g-factor\cite{usher1990,sondhi1993,Barrett1995,Schmeller1995,Maude1996,shukla2000}.
Another consequence of this energy cost is that the charged
excitations at $\nu=1$ have a non-trivial long-range spin order,
with a slow rotation of the spin along the radial direction
\cite{sondhi1993}. The size ($s$) of such a spin texture, or
"Skyrmion", is the number of electrons participating in the
excitation, and depends on the ratio of the Coulomb energy to the
Zeeman energy
\cite{sondhi1993,Barrett1995,Schmeller1995,Maude1996,shukla2000}.
The larger the Zeeman energy, the less preferable it is for the
Skyrmion to form, and $s$ tends to one as the g-factor of the 2DES
is sufficiently increased. While in the context of a 2DES a
Skyrmion is usually associated with the spin degree of freedom,
these excitations should exist in any system whose Hamiltonian has
an SU(2) symmetry \cite{sondhi1993,arovas1999}.

In a 2DES with two equivalent and energy degenerate conduction
band valleys, there is a direct analogy between the valley index
(isospin) and the electron spin. Indeed, it has been argued that,
since rotations in the valley isospin space leave the electron
Hamiltonian unchanged, a 2DES with a valley degree of freedom
contains a hidden SU(2) symmetry \cite{rasolt1990}. In a
two-valley system, an externally applied strain has the same
effect on the valleys as the magnetic field has on the electron
spins; the strain breaks the isospin SU(2) symmetry by lifting the
energy degeneracy between the valleys. In this Letter, we directly
probe the effects of interaction on a 2DES confined to an AlAs
quantum well where the electrons occupy two valleys. Via the
application of symmetry breaking in-plane strain, we tune the
energies of these valleys and measure the $\nu=1$ QHS energy gap.
The gap's dependence on strain provides experimental evidence that
the excitations of this QHS are valley Skyrmions (valley
textures).

AlAs is an indirect gap semiconductor in which the electrons
occupy multiple valleys at the X-point of the Brillouin zone. The
constant energy surface in bulk AlAs consists of three highly
anisotropic ellipsoids (six half-ellipsoids) with their major axes
along the $<$100$>$ crystal directions; we designate these valleys
by their major axis' direction. The longitudinal and transverse
effective masses of the electrons in these valleys are 1 and 0.2,
respectively, in units of the free electron mass. While the
valleys in bulk AlAs are energy degenerate, this degeneracy is
lifted when the electrons are confined to a 2D layer. In a quantum
well grown on top of a GaAs (001) wafer, only the [100] and [010]
valleys are occupied for well widths greater than $\sim$ 5 nm
\cite{maezawa1992}. This is different from the Si (001) MOSFET
(metal oxide semiconductor field-effect transistor), in which the
two valleys with the out-of-plane major axes are occupied.

We study 2D electrons in an 11 nm-wide, modulation doped AlAs
quantum well grown on a GaAs (001) substrate using molecular beam
epitaxy \cite{depoortere2002}. The electrons are confined to AlAs
by two Al$_{0.4}$Ga$_{0.6}$As barrier layers. We inject current
into the 2DES through alloyed AuGeNi contact pads and measure the
longitudinal resistance on an etched Hall bar mesa aligned with
[100]. The sample is cooled in a pumped $^3$He system that allows
us to perform measurements in the temperature ($T$) range between
0.3 K and 6 K. The typical mobility of our samples is between 8
and 15 m$^2$/Vs for the electron concentration
$n=2.6\times10^{11}$ cm$^{-2}$ at $T=0.3$ K \cite{mobility}.
\begin{figure*}
\includegraphics[scale=.8]{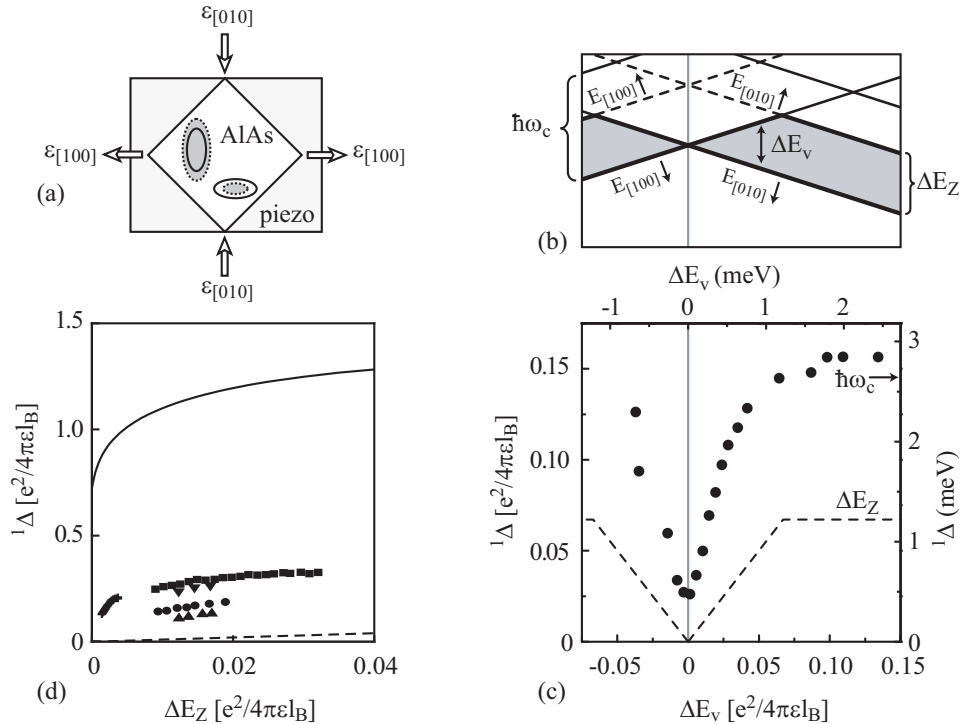}
\caption{(a) In-plane strain breaks the valley energy degeneracy;
e.g., tension along [100] results in a transfer of electrons from
the [100] valley to the [010] valley. (b) Single-particle fan
diagram for the AlAs 2DES using band parameters. (c)
Single-particle (dashed lines) and measured (circles) $\nu=1$ QHS
gap, $^1\Delta$, as a function of the single-particle valley
splitting, $\Delta E_v$, for $n=2.55\times10^{11}$ cm$^{-2}$.
$^1\Delta$ and $\Delta E_v$ are given both in meV (top and right
axes) and also in units of the Coulomb energy $e^2/4\pi\epsilon
l_B$, where $\epsilon=10\epsilon_0$ is the dielectric constant of
AlAs and $l_B=(\hbar/eB)^{1/2}$ is the magnetic length; in our
experiments, $B=10.5$ T and $e^2/4\pi\epsilon l_B=18.2$ meV. (d)
$^1\Delta$ vs. $\Delta E_Z$ for spin Skyrmions in a single-valley
2DES. The dashed line is the single-particle Zeeman splitting. The
theoretical prediction for the Skyrmion excitation energy is shown
by the solid curve and the filled symbols show the experimentally
determined gaps at $\nu=1$ (see text).} \label{valleysplitting}
\end{figure*}

To control the energy splitting between the valleys, we strain the
sample by gluing it to one side of a PZT-5H piezoelectric stack
(piezo) with the sample's [100] direction aligned with the poling
direction of the piezo, as shown in Fig. \ref{valleysplitting}
(a). The voltage-induced strain change of the piezo allows us to
\textit{in situ} and continuously vary the in-plane strain applied
to the sample \cite{shayegan2003} and thereby lift the valley
degeneracy. As shown schematically in Fig. \ref{valleysplitting}
(a), by applying tension along the [100] direction, we induce a
transfer of electrons from the [100] valley into the [010] valley;
the \textit{total} density of the 2DES, however, remains costant
to better than 1\%. The \textit{single-particle} valley splitting,
$\Delta E_v$, can be expressed in the form $E_2\epsilon_{xy}$,
where $E_2=5.8$ eV is the shear deformation potential for AlAs
\cite{charbonneau1991} and $\epsilon_{xy}$ is the difference in
the strain along [100] and [010]. We determine $\Delta E_v$ to
within an accuracy of 0.1 meV from the oscillations of the
sample's resistance at high integer filling factors as a function
of strain \cite{shayegan2003} and corroborate it with the Fourier
analysis of Shubnikov-de Haas oscillations, the zero-field
piezoresistance data \cite{shkolnikov2004b}, and our calibration
of the piezo-induced strain \cite{shayegan2003}.

Before presenting the experimental data, it is instructive to
describe the fan diagram of the AlAs 2DES as a function of
strain-induced valley splitting in a \emph{non-interacting},
\emph{single-particle} picture [Fig. \ref{valleysplitting} (b)].
The magnetic field perpendicular to the plane of the 2DES
quantizes the orbital motion of the electrons and forces them to
occupy a discrete set of energy levels separated by the cyclotron
energy, $\hbar\omega_c=\hbar eB/m^*$, where $m^*=0.46$ is the
cyclotron effective mass in AlAs \cite{lay1993}. There are four
sets of these Landau levels, one for each spin and valley
combination. The energy splitting between oppositely polarized
spins is controlled through $\Delta E_Z=g_b\mu_B B$ (band g-factor
$g_b=2$), while the levels corresponding to different valleys are
separated by $\Delta E_v$. When $\epsilon_{xy}=0$, in the
single-particle picture of Fig. \ref{valleysplitting} (b), the
energy degeneracy of the Landau levels associated with different
valleys implies that there is no $\nu=1$ QHS. As the applied
strain increases and breaks the valley degeneracy, the $\nu=1$ QHS
gap ($^1\Delta$) develops. For $\Delta E_v<\Delta E_Z$, $^1\Delta$
should increase linearly with strain and be equal to $\Delta E_v$.
For sufficiently large strains such that $\Delta E_v$ exceeds
$\Delta E_Z$, $^1\Delta$ should be equal to $\Delta E_Z$,
independent of strain.

The results of our measurements are in sharp contrast to the
simple, non-interacting picture described above.  Experimentally,
we determine $^1\Delta$ from the activated T-dependence of the
longitudinal resistance, $R_{xx}$, according to $R_{xx} \sim
\exp(-^1\Delta/2k_B T)$. This gap is plotted in Fig.
\ref{valleysplitting} (c) as a function of $\Delta E_v$. The
measured $^1\Delta$ differs from the single-particle picture
[dashed curve in Fig. 1(c)] in three important aspects;
$^1\Delta$: (1) is finite even when there is no applied strain,
(2) rises with the applied strain much faster than expected, and
(3) saturates at a value which is closer to $\hbar\omega_c$ than
to $\Delta E_Z$. As we discuss in the remainder of the paper,
these features of the data are all consistent with the many-body
origin of $^1\Delta$ and the presence of valley Skyrmions at small
values of strain \cite{silicon2}.

Since there are no explicit calculations for the dependence of
$^1\Delta$ on $\Delta E_v$, we compare our data to the dependence
of $^1\Delta$ on $\Delta E_Z$ in a system which has a spin degree
of freedom but no valley degree of freedom
\cite{Schmeller1995,shukla2000}. In Fig. \ref{valleysplitting}
(d), we show the theoretical prediction for $^1\Delta$ (solid
line) that includes the effects of electron-electron interaction
\cite{shukla2000}. Even when $\Delta E_Z = 0$, $^1\Delta$ is
nonzero and has a value of 0.63 in units of the Coulomb energy.
This is an interaction-induced gap, and rises rapidly as a
function of $\Delta E_Z$, with the slope of the $^1\Delta$ vs.
$\Delta E_Z$ curve equal to the Skyrmion size $s$. The qualitative
agreement between the theoretical predictions and the measured
$^1\Delta$ in GaAs and AlGaAs 2DESs [filled symbols in Fig.
\ref{valleysplitting} (d)] are widely interpreted as evidence for
the existence of QHS spin Skyrmions
\cite{Schmeller1995,Maude1996,shukla2000}. The striking
resemblance between the data of Figs. 1(c) and (d) suggests the
presence of QHS "valley Skyrmions" in AlAs 2DES. With this
interpretation, the value of $s$ for the valley Skyrmions in our
2DES is 4.3 at $\Delta E_v/(e^2/4\pi\epsilon l_B)=0.01$,
comparable to $s=5$ for the spin Skyrmions in GaAs 2DESs at
$\Delta E_Z/(e^2/4\pi\epsilon l_B)=0.01$ \cite{Schmeller1995}.

In our experiments, as in previous measurements of spin Skyrmions
in single-valley systems
\cite{Schmeller1995,Maude1996,shukla2000}, we find that the
measured values for $^1\Delta$ are only a small fraction of the
Coulomb energy, in sharp contrast to the theoretical prediction
[see Fig. 1(d)]. The strongly depressed experimental gap for
$\nu=1$ in a non-idealized 2DES is attributed to the combination
of the effects of the finite wavefunction thickness along the
confinement direction, Landau level mixing, and disorder
\cite{shukla2000,sinova2002}. In AlAs, the effect of Landau level
mixing should be particularly important since the Coulomb energy
in our experiment is much larger than the separation between the
single-particle energy levels.
\begin{figure}
\includegraphics[scale=0.8]{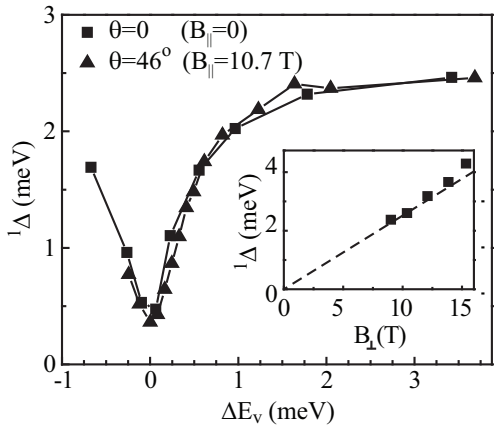}
\caption{Dependence of $^1\Delta$ on $\Delta E_v$ for an AlAs 2DES
with $n=2.55\times10^{11}$ cm$^{-2}$ at different tilt angles.
Single-particle $\Delta E_Z$ at $\nu=1$ is 1.22 meV and 1.75 meV
for $\theta=0$ and $\theta=46^o$, respectively. Inset shows a
comparison between $^1\Delta$ (squares) and $\hbar\omega_c$
(dashed line) measured at $\theta=0$ as a function of
perpendicular magnetic field (density) for a 2DES with only the
[010] valley occupied ($\Delta E_v=3.4$ meV).} \label{tilted}
\end{figure}

We now address whether the spin degree of freedom affects the
strength of the $\nu=1$ QHS in our samples at both large and small
values of $\Delta E_v$. In Fig. 1(c) we note that, for large
values of $\Delta E_v$, where the electrons occupy only the [010]
valley, the measured $^1\Delta$ saturates at a value that is close
to $\hbar\omega_c$ rather than $\Delta E_Z$, contrary to what we
would expect from the fan diagram of Fig. 1(b). We have measured
this saturation value as a function of the 2DES density (Fig. 2
inset), and we find that it is indeed always close to
$\hbar\omega_c$ in our accessible density range. This observation
is not surprising. It is consistent with previous measurements
\cite{papadakis1999,shkolnikov2004} which indicate that, thanks to
interaction, $\Delta E_Z$ is greatly enhanced in AlAs 2DESs and is
indeed larger than $\hbar\omega_c$ for the parameters of our
sample. Therefore, in the limit of very large $\Delta E_v$, the
nearest single-particle energy level is $\hbar\omega_c$ above the
ground state energy. Our data indicate that once $^1\Delta$
reaches $\hbar\omega_c$, it remains fixed at $\hbar\omega_c$ and
no longer shows an enhancement over the single-particle gap.

To further understand the role of electron spin in our 2DES, and
to ensure that our data of Fig. 1(c) are not influenced by the
spin degree of freedom, we performed the following experiment.  We
measured $^1\Delta$ as we increased the Zeeman energy by tilting
the normal of the 2DES by $\theta$ degrees with respect to the
magnetic field while keeping the perpendicular component of the
field $B_\bot$ (and thus $\hbar\omega_c$) constant \cite{homegac}.
In Fig. \ref{tilted}, we show the measured $^1\Delta$ for
$\theta=0$ (squares) and $\theta=46^o$ (triangles). We display the
data at $\theta=46^o$ shifted by +0.27 meV in $\Delta E_v$ with
respect to the $\theta=0^o$ data \cite{tilt}. As can be clearly
seen, $^1\Delta$ in unaffected by a $43\%$ increase of $\Delta
E_Z$. These data thus indicate that the dependence of $^1\Delta$
on $\Delta E_v$ seen at small values of strain in Fig. 1(c) is
primarily a result of the valley interaction and not spin effects.

Our $R_{xx}$ vs. magnetic field data (Fig. 3), taken at various
values of strain, reveal additional interesting features.  The low
temperature ($T$ = 0.3 K) traces shown in the main figure exhibit
several strong QHSs at various fillings, including one at the
fractional filling $\nu=2/3$. The strongest QHS clearly occurs at
$\nu=1$ in the entire range of applied strains, attesting to the
robustness of this QHS even in the limit of zero strain. In traces
taken at higher temperature ($T$ = 2 K), plotted in the inset, the
$\nu=1$ minimum is indeed the most prominent feature of the data.
The $T$ = 2K data show a weakening of the $\nu=1$ minimum when the
strain is small, consistent with $^1\Delta$ values of Fig. 1(c).
More remarkably, we observe that for small strains there are in
fact two minima near $\nu=1$, one exactly at $\nu=1$ and the other
at a slightly lower field. The side minimum, whose resistance
value does not follow a simple, thermally activated behavior,
disappears with increasing magnitude of $\Delta E_v$. The
appearance of a double minimum near a QHS has been previously
associated with a phase transition between two competing QHSs,
such as between spin-polarized and unpolarized QHSs at $\nu=8/5$
\cite{eisenstein1989} or at $\nu=2/3$ \cite{engel1992} in GaAs
2DESs. We do not know if the double minimum we observe is a
signature of a phase transition in the AlAs 2DES; we note however,
that competing states in 2DESs with (spin) SU(2) symmetry have
been theoretically suggested \cite{sinova2002}.
\begin{figure}
\includegraphics[scale=.85]{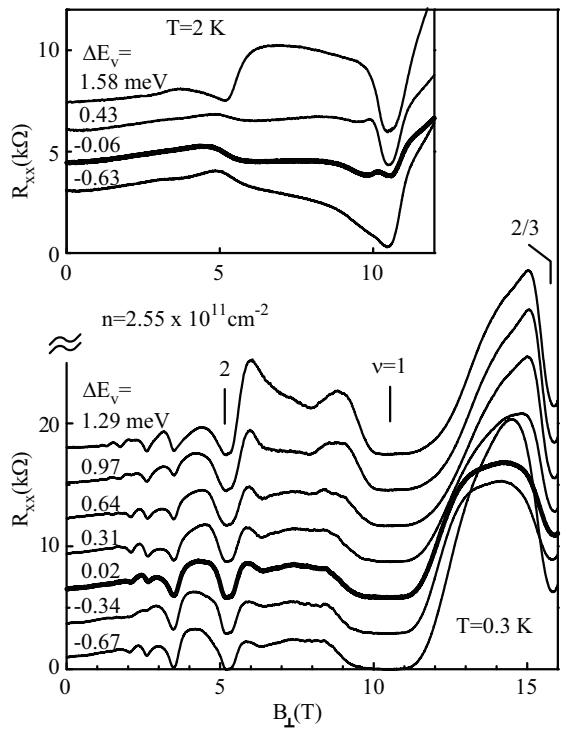}
\caption{Magnetoresistance of the AlAs 2DES for different values
of applied strain at $T$ = 0.3 K (main figure) and 2 K (inset).
The value of $\Delta E_v$ is indicated for each trace; the bold
traces correspond to the smallest magnitude of $\Delta E_v$.
Traces are vertically offset from each other for clarity.}
\label{waterfall}
\end{figure}

In conclusion, we find that electron-electron interaction strongly
modifies the excitations of the $\nu=1$ QHS of a two-valley 2DES.
Our measurements of the energy gap for this QHS in an AlAs 2DES as
a function of applied strain suggest that these excitations
involve valley Skyrmions. In closing, we note that while nuclear
magnetic resonance measurements serve as a powerful probe of spin
Skyrmions \cite{Barrett1995}, they cannot be used to detect valley
Skyrmions. It may be possible, however, to use electromagnetic
radiation to couple to the valley Skyrmions, as rotations in the
valley isospin space should result in a change in the conductivity
tensor of the AlAs 2DES.

\begin{acknowledgments}
We thank the NSF for support, and E. Tutuc, S. Sondhi, A. H.
MacDonald, and D. P. Arovas for illuminating discussions.
\end{acknowledgments}

\begin{thebibliography}{23}
\bibitem[{mob()}]{mobility}
\bibinfo{note}{Since each valley has a different and anisotropic effective
  mass, the mobility of our samples is a function of the relative valley
  occupancy and thus the applied strain.}
\bibitem[{hom()}]{homegac}
\bibinfo{note}{Data of Fig. 2 were taken during a different sample cooldown.
  The different state of the 2DES is likely responsible for the slightly
  different values of $^1\Delta$ in Fig. 1 (c) and Fig. 2.}
\bibitem[{til()}]{tilt}
\bibinfo{note}{The shift of $^1\Delta$ is likely a result of the finite thickness of the electron layer and the ensuing
parallel-field induced change in the confinement energy of the two
valleys.}
\end{thebibliography}

\end{document}